\documentstyle[amsfonts,preprint,aps]{revtex}

\begin{document}
\draft
\preprint{ }

\title{
 Exact spectra, spin susceptibilities and order
 parameter of the quantum Heisenberg
 antiferromagnet on the triangular lattice.
}

\author{
B. Bernu\thanks{Laboratoire de Physique Th\'eorique des
    Liquides, Universit\'e P. et M. Curie, boite 121, 4 Place Jussieu, 75252
    Paris Cedex. URA 765 of CNRS},
P. Lecheminant $ ^* $\thanks{Groupe
    de Physique Statistique, Universit\'e de Cergy-Pontoise,
    95806 Cergy-Pontoise Cedex.},
C. Lhuillier $ ^* $, and
L. Pierre\thanks{
    U.F.R. SEGMI, Universit\'e Paris-X, Nanterre, 92001 Nanterre Cedex.}\\
}
\date{\today}
\maketitle

 \bibliographystyle{prsty}

\begin{abstract}
    Exact spectra of periodic samples are computed up to $ N=36 $.
    Evidence of an extensive set of low lying levels,
    lower than the softest magnons, is exhibited.
    These low lying quantum states are degenerated in
    the thermodynamic limit;
    their symmetries and dynamics as well as their finite-size
    scaling are strong arguments in favor
    of N\'eel order.
    It is shown that the N\'eel order parameter agrees with
    first-order spin-wave calculations.
    A simple
    explanation of the low energy dynamics is given as well as
    the numerical determinations of the energies, order parameter
    and spin susceptibilities of the studied samples. It is shown how
    suitable boundary conditions, which do not frustrate N\'eel
    order, allow the study of samples with $ N=3p+1 $ spins.
    A thorough study of these situations is done in
    parallel with the more conventional case $ N=3p $.
\end{abstract}

\newpage
\narrowtext

\section{Introduction}
	The nature of the thermodynamic ground-state of the spin-$ 1/2 $
Heisenberg antiferromagnetic hamiltonian, in two space dimensions,
 is still an open question.
There have been considerable amounts of theoretical or numerical works
on Heisenberg antiferromagnets[1-45] but few exacts results.
It is known that in one and two dimensions the system is disordered at
 $ T \neq 0 $\cite{mw66}, and that the one-dimensional
system does not exhibit N\'eel order even at $ T=0 $.
The two-dimensional case is more controversial. There is a rather large
consensus on the existence of  N\'eel order at $ T=0 $ on the (unfrustrated)
square lattice\cite{a52,lda88,h88,ry88,he88,th89,gsvs89,tc90,l90,c89,sz92}.
The situation is much more puzzling as regards the  triangular lattice case.
It was indeed the first system to be proposed by Anderson\cite{a73}
and Fazekas\cite{fa74} as a candidate for a spin-liquid.
On this lattice the ``frustration'' implies that the classical system is
not very stable ($ E_{cl} = < 2 {\bf s}_{i}.{\bf s}_{j}> = -1/4 $), and
the spin-wave calculations predict an important reduction (by about one
half) of the sublattice magnetization by quantum
fluctuations\cite{nm85,jlg89,m92}.
Perturbation theory\cite{okn87}, series expansions\cite{sh92}
and high temperature calculations\cite{esy93} have been
developed which suggest that the spin-wave
calculations possibly underestimate this renormalization.
Many variational calculations have been done exhibiting either ordered
\cite{he88,sll94} or disordered solutions\cite{ont86,kl87,ywg93}.

	In the square lattice case, numerical methods (Q.M.C., Ulam's
or Trotter-Suzuki methods) have brought very interesting indications
on N\'eel order\cite{ry88,gsvs89,tc90,l90,c89}. Unfortunately these
methods which allow to handle large samples cannot be applied
to the triangular case: they lay
on a property of positivity of offdiagonal matrix elements which is
violated in the triangular case; it is the well known sign problem
which plagues many studies of strongly correlated fermions.
Exact diagonalizations of the hamiltonian are thus the last resort to
gather new information on these models. This approach has been
developed by other authors\cite{mb79,f87,nn88,i87,i89,jdgb90,demw90,lr93}.
Most of them conclude to the absence of N\'eel order
for the triangular Heisenberg antiferromagnet (THA)  at $ T=0 $.
But two requirements are to be met to analyze the raw numerical data:
a consistent finite-size scaling analysis and a quantum definition of
observables.
With these two constraints we will show in this paper that all
the numerical data point to an ordered ground-state for the THA.

As the second and perhaps more important objective we want to illustrate
the spectral properties of an ordered quantum antiferromagnet on
finite lattices
and how they embody the characteristics of the symmetry breaking state
(parts of these results have already been published\cite{blp92}).

In section II, we first show the numerical spectra of the THA on periodic
samples with N=9 up to N=36 spins and we present
the essential characteristics of these spectra. Their low lying levels
form two families: the first one contains an extensive number
of  states called QDJS in
ref\cite{blp92}, which collapse to the ground-state as $ N^{-1}$, and
are characterized by the spatial symmetries of the classical
N\'eel ground-state. The second family collapse more slowly
to the ground-state as $ N^{-1/2}$  and forms the familiar one
magnon excitations. Such a structure has been conjectured
a long time ago by Anderson
in his seminal paper on antiferromagnets\cite{a52}, and the subject
has already been studied in the square lattice case
\cite{th89,gsvs89,gsvs289,khl89,nz89,f89}.

Section III is devoted to the understanding of the
QDJS of the triangular Heisenberg hamiltonian. We show
how the hypothesis of a N\'eel order explains the number of
levels appearing for each spin, their symmetries and their
dynamics. The analysis of these spectra gives a first information
on the spin susceptibilities $\chi_{\perp}$ and  $\chi_{\parallel}$
of the THA. In order
to check our general ideas, we extend the analysis to samples
with  $ N=3p+1 $ spins. Suitable boundary conditions allow us to study
situations which do not frustrate N\'eel order. We compare
all the results of energies and susceptibilities with the first-order
spin-wave results.

In section IV, we analyze the static spin-spin structure function and
the N\'eel order parameter.
We show how diagonalization results sustain the hypothesis of N\'eel
order and how the analysis of some previous authors was in error,
leading to confusing conclusions.

We conclude by a
brief discussion of the interest and limitations of this approach.

In the Appendices, we develop the group theoretical analysis of the THA
and explain our numerical method: this is a technical but crucial point
in order to obtain the needed information in a minimum computer
time, within memory capacities of today computers. We had
for example to make sure that the low lying levels of all the different
irreducible representations order as they should in the framework of
our hypothesis. This leads us to compute the whole spectrum
of the samples up to $ N=21 $ and a large number of low lying levels
in each irreducible representations for all sample sizes up to
 $ N=27 $.

\section{Spectra of periodic samples: an overview}

All the results presented here have been obtained by diagonalization of
the spin-$ 1/2 $ Heisenberg hamiltonian of periodic samples on the
triangular lattice (\cite{mb79} and Appendix A for more details).
The Heisenberg hamiltonian reads:
   \begin{equation}
      \label{eq-heis}
      {\cal H} = \sum_{<i,j>} 2\;\; {\bf s}_i.{\bf s}_j
   \end{equation}
   where the sum runs over pairs of nearest neighbors and
   ${\bf s}_i $, ${\bf s}_j$ are the spin-$1/2$ operators
   on sites $i$ and $j$.

In this section, we are concerned with periodic samples with $ N=3p $.
Such samples do not frustrate
the classical N\'eel order.
A classical  N\'eel state on the triangular lattice has
coplanar spins with a three-fold
rotational symmetry defining three magnetic sublattices ($A,B,C$)
(see fig.\ref{fig:classical}).
On each sublattice, the spins are ferromagnetically aligned and the
angles between the magnetizations of two sublattices are $\pm  2 \pi
/3 $.
The total spin of a triangular plaquette is zero and the rotations of
the spins around an $ABC$ triangle could be clockwise or counter-clockwise
defining two opposite helicities.
The point symmetry group of these classical solutions is $C_{3v}$, and
the translational symmetry group is that of a sublattice, say $A$.
These classical
solutions break the translation and spin rotation invariance of
Eq.\ref{eq-heis} whereas the quantum eigenstates that we
will now consider do not.

Using group theory, described in Appendix A, we have
computed the complete spectrum of Eq.\ref{eq-heis} for
 $N=9,12,21$.
The 5 lowest energies have been computed for $N=27$ in all the
   irreducible representations (IR), and the 3 lowest energies for
   $N=36$ in the homogeneous states $\bf{k}=\bf{0}$, invariant under
   rotation.
   The energy spectra are given in fig.\ref{fig:spectra} and the
   lowest levels listed in table \ref{table-lowest-energies}.
   Our results are in perfect agreement with previous
   diagonalization results\cite{nn88,lr93}.

A first very striking feature of these spectra could be read in
fig.\ref{fig:spectra}: the lowest energy levels in each spectra order
   with increasing $S$.
   This is strongly reminiscent of the theorem by Lieb and Mattis for
   bipartite lattices\cite{lm62}. But
   there is presently no proof that the property holds for other
lattices. We find this is true for other non bipartite problems
like the Kagom\'e Heisenberg antiferromagnet or for the $ J_{1}-J_{2}$
models on triangular and Kagom\'e lattices.

The second striking feature of these spectra is the existence in
each $S$ subspace of a family of low lying levels well separated
from the others:
we called these
   states the ``Quasi-Degenerate-Joint States'' (QDJS) in ref.
\cite{blp92}.
In fig.\ref{fig:spectra}, it is shown that
   the QDJS energies ($6N<{\bf s}_{i}.{\bf s}_{j}>$) stand
   around a line $E=S(S+1)/(2I_N)$, where the moment of inertia $I_N$ is
essentially proportional to $N$ (see fig.\ref{fig:inertia}).
This suggests that these levels collapse as $ N^{-1}$ to the
ground-state.
We shall show in the next section that this family of ${\cal O}(N^3) $
QDJS has all the properties expected for the description of N\'eel
quantum ground-states: in particular, the $ C_{3v}$ and magnetic
sublattice translation invariance of the N\'eel states
(they only contain the ${\bf k} = {\bf 0}$ or $\pm {\bf k}_{0}$  IRs
of the lattice translation operators where
 $\pm {\bf k}_{0}$ are the two wavevectors of the corners of the crystal
Brillouin zone, mapping on the center of the magnetic Brillouin zone).
The symmetry breaking N\'eel states are linear combinations of these
QDJS.

To study higher excited states, let us first come back to the
 $N=9$ spectrum of fig.\ref{fig:transitions}.
Above the basal line of QDJS one sees very clearly
 two families of levels.

The first excited-state family (horizontal bars in
fig.\ref{fig:transitions}) consists of eigenstates of the translation
operator with ${\bf k}\ne{\bf 0}$ and ${\bf k} \ne \pm{\bf k}_0$,
they are typically states involved in one magnon excitations.
What is usually called a magnon in solid state physics is in fact a
spatial modulation of the N\'eel state.
As the N\'eel states should be seen  as linear combinations of the QDJS,
the one magnon excitations are in fact linear superpositions of this
first family of excited states.
A one magnon excitation is thus pictured
in the spectrum of fig.\ref{fig:transitions} as a $\Delta S = 1 $
collective excitation of the QDJS (three legs symbols family) towards
the $k$ family of levels (this picture has been ascertained by
the computation of the dynamic structure factor of the first QDJS
of the $ N=12 $ sample\cite{bllp93}).

The second family of excited states (black triangles)
 belongs to the non trivial IRs
of $C_{3}$ [${\cal R}_{\frac{2 \pi}{3}} \Psi =
\exp (\pm i 2 \pi /3) \Psi$], this is the first family of levels
(labeled $3-$ in the following)
that could sustain chiral states (${\cal R}_{2\pi/3}$ is a spatial rotation
of angle $2\pi/3$).
The last family of excited states (open triangles)
 belongs to the non trivial IR of $C_{3v}$,
the states invariant under a rotation of $2\pi/3$
and odd under axial symmetry ($\sigma_x$)
[${\cal R}_{\frac{2 \pi}{3}} \Psi = \Psi, \sigma_{x}\Psi =-\Psi$]
(labeled $3=~x-$, see table III).

For larger $N$, the spectra become more and more dense, but this
hierarchy of levels is always obeyed.
The magnon states collapse to the ground-state roughly as $k$ i.e.
as $N^{-1/2}$ whereas the energy of the chiral states seems to have a
gap ($E_{3-} - E_{0} = $
6, 6, 1.908, 2.958, 0.7661, 2.454, 2.641, 2.01, 2.546,
for $N=$ 7, 9, 12, 13, 16, 19, 21, 25, 27).
The hypothesis of a chiral ground-state on the triangular lattice seems
excluded by our results\cite{b89,t92}.
The energies of the $3=~x-$ states are much higher in the spectra
(out of the range of energies of table I).

\section {Qualitative analysis of the quasi-degenerate ground-state
multiplicity}

The symmetries and the dynamics of the QDJS are essential to understand
the nature of the order in the thermodynamic limit.
Let us begin our investigation by the symmetry analysis of the
eigenspectrum.

\subsection{The QDJS as the coupling of three spins}

It is straightforward to verify (see for example Table
\ref{table-lowest-energies}) that the QDJS family exhibits the exact
number of states expected from the coupling of three spins of length
 $ N/6 $:
\begin{equation}
   N_{S}= \min(2S+1,N/2-S+1)
\label{eq-muls}
\end{equation}
 This number is readily obtained  by noticing that the Hilbert space
of two spins
 $ S_{A}=S_{B}=N/6 $ can be split in $ (N/3+1) $ subspaces associated
to the eigenvalues of their sum $ S_{A+B}$
(with $ |S_{A}-S_{B}|=0  \leq S_{A+B} \leq S_{A}+S_{B}= N/3$).
The coupling of $ S_{A+B}$ with $ S_{C}$ gives then a total spin S
($|S_{A+B}-N/6| \leq S \leq S_{A+B}+ N/6$), which leads to result
Eq.\ref{eq-muls}.

Such a property is not unexpected if we look at that
family of levels as arising from the renormalization of the
classical N\'eel state by quantum fluctuations.
In this purpose it is most useful
to Fourier analyze the hamiltonian Eq.\ref{eq-heis}; separating the
 ${\bf k} =  {\bf 0}$ and $\pm {\bf k}_0 $ contributions from the others,
Eq.\ref{eq-heis} reads:
 \begin{equation}
{\cal H}  = {\cal H}_{0}+{\cal V}
 \end{equation}
where
 \begin{equation}
{\cal H}_{0} = \frac{9}{N} \left({\bf S}^{2} - {\bf S}_{A}^{2}
- {\bf S}_{B}^{2}-  {\bf S}_{C}^{2}\right)
\label{eq-a-8}
 \end{equation}
 ${\bf S}_A$ (resp. ${\bf S}_B, {\bf S}_C) $ is the total spin of the $ A $
 (resp. $ B, C) $ sublattice, and
 \begin{equation}
{\cal V} = \sum_{k \neq 0;\pm k_{0}} f({\bf k}) \bf S_{k}. \bf S_{-k}
\label{eq-a-9}
\end{equation}
with
\begin{equation}
      {\bf  S_{k}} =  \frac{1}{\sqrt{N}} \sum_{i} {\bf  s}_{i}
                                 \exp(i {\bf k} .{\bf r}_{i})
\end{equation}
and
\begin{equation}
 f({\bf k}) = \exp(i {\bf k}.{\bf u}_1) +
\exp(i {\bf k} .{\bf u}_2) + \exp( - i {\bf k} .({\bf u}_1
+{\bf u}_2)) +  c.c
\label{eq-gam}
 \end{equation}
where ${\bf u}_1$ and ${\bf u}_2$ are the two basis vectors of the lattice.

 ${\cal H}_{0}$ commutes with ${\bf S}_A, {\bf S}_A^2$,
 ${\bf S}_B, {\bf S}_B^2$, ${\bf S}_C, {\bf S}_C^2$
and the eigenstates of ${\cal H}_{0}$  (written as
 $| \Psi_0(i, S, M_{S}) >$)
are eigenstates of  ${\bf S}^2$, ${\bf S}_A^2$,
 ${\bf S}_B^2$, ${\bf S}_C^2$
with eigenvalues:
\begin{equation}
E(S, S_{A}, S_{B}, S_{C}) = \frac{9}{N} \left(S(S+1) -  S_{A}( S_{A} + 1)
 -  S_{B}( S_{B} + 1) -  S_{C}( S_{C} + 1)\right).
 \end{equation}
For each $S$ value the lowest eigenstates
of ${\cal H}_{0}$ are
obtained for
 $ S_{A}= S_{B}= S_{C} =  \frac{N}{6}$,
their energies are:
\begin{equation}
E_{0}(S) = - \frac{3}{4} (N+6) + \frac{9}{N} S(S+1).
\label{eq-a-11}
\end{equation}
These states $|\Psi_0^0(i, S, M_S)>$ (with $i$ from 1 to
 $N_{s}$), fully polarized on each
magnetic sublattice, are the projections of the classical N\'eel states on
the various IRs of $SU(2)$. They present the usual $2S+1$ degeneracy
associated to $M_{S}$ multiplied
by the number $N_{S}$ of different couplings of three spins $N/6$
(Eq.\ref{eq-muls}).

As ${\bf S}_A^2 $,
 ${\bf S}_B^2 $,  ${\bf S}_C^2 $
 do not commute with $\cal V $,
 $ | \Psi_0^0(i, S, M_S) > $ are not eigenstates of $\cal H $, but
 we can look at them
as the first (bad) approximation to the exact QDJS:
the perturbation $\cal V $ dresses these ``classical states'' with
quantum fluctuations,
 decreasing the average value
of the sublattice magnetizations
and lowering their energy towards the exact results. This process can
lift the $ N_{S}$ degeneracy of the  $ | \Psi_0^0(i, S, M_{S}) > $ and
it is indeed what is observed in the exact spectra of QDJS, but we verify in
 fig.\ref{fig:spectra} and Table
\ref{table-lowest-energies}
that the set of QDJS has exactly the correct
multiplicity $ N_{S}$.

\subsection{Symmetries of the QDJS }

The QDJ eigenstates belong to the three following IRs of
the space symmetry group:

 $\Gamma_1$ : [ ${\bf k} = {\bf 0}$, ${\cal R}_\pi\Psi=\Psi$
                                    , ${\cal R}_{2\pi/3}\Psi=\Psi$
                                    , $\sigma_x\Psi=\Psi$ ],
 $\Gamma_2$ : [ ${\bf k} = {\bf 0}$, ${\cal R}_\pi\Psi=-\Psi$
                                    , ${\cal R}_{2\pi/3}\Psi=\Psi$
                                    , $\sigma_x\Psi=\Psi$ ]
and
 $\Gamma_3$ : [ ${\bf k} = \pm {\bf k}_{0}$
                                    , ${\cal R}_{2\pi/3}\Psi=\Psi$
                                    , $\sigma_x\Psi=\Psi$ ],
where ${\cal R}_\phi$ is a rotation of angle
 $\phi$ and $\sigma_x$ is an axial symmetry.

They appear with regular rules (described below in III.D and
in Appendix B) in all the IRs of SU(2): that is for each $S$ value.

This proves that the QDJ eigenstates are invariant:

 $ 1) $ under translations of the magnetic sublattices. They
only contain the ${\bf k} = {\bf 0}$ or $\pm {\bf k}_{0}$ IRs
of the lattice translation operators.

 $ 2) $ under the point group $ C_{3v}$ (${\cal R}_{2\pi/3}\Psi=\Psi $,
 $\sigma_x\Psi=\Psi $).

The appearance of the ${\bf k} = {\bf 0}$ and $\pm {\bf k}_{0}$ IRs of
the translation group in this quasi-degenerate ground-state multiplicity
allows to build states which breaks the translational symmetry of the
lattice (as it is the case of the N\'eel state).
In the same approach, the appearance of all the IRs of  $ C_{2}$ (states
where  ${\cal R}_\pi\Psi=\pm \Psi$) allows to break the inversion symmetry,
whereas the presence of all IRs of $ SU(2) $ allows to build states with a
vectorial magnetization pointing in a given direction.
In summary, all the IRs which keep invariant the N\'eel states
appear in the QDJS and no others.
Note that the low lying levels of ${\cal H}_{0}$ described in the
previous subsection belong to the same IRs than the QDJS.
It is straightforward to verify that the quantum perturbation ${\cal V}$
is also invariant under the symmetry group of the classical N\'eel
states; thus the renormalization  by ${\cal V}$ of the sublattices
spins can take place without disturbance of the symmetries of the classical
N\'eel states.

The picture of the N\'eel states as linear combinations of the QDJS
becomes plausible.
But the symmetry argument does not prove by itself that the low lying
levels of ${\cal H}_{0}$ have evolved identically in the dressing by
the quantum fluctuations induced by ${\cal V}$.
For example, the simple question ``do all the QDJS have the same
extensive sublattice magnetization?'' cannot be answered from only
symmetry arguments.
A still more basic issue is discussed in the following subsection.

\subsection {The effective dynamics of the QDJS}

We have shown in the previous subsections that the QDJS possess
 the same space symmetries than the
(sphericalized) classical N\'eel states.
We have now to make sure that the dynamics of this set of states
can be reduced in the thermodynamic limit to that of a collective
variable: the order parameter of the antiferromagnet. This is
indeed a very serious issue and the necessary condition for the
rigidity of the supposed-to-be ordered phase.

To precise the nature of this order parameter, let us concentrate on
a presumed  N\'eel order on the triangular lattice. Defining
a specific N\'eel state requires the knowledge of exactly
three angles: two angles locate
the helicity ${\bf \Upsilon }$ defined as:
\begin{equation}
{\bf \Upsilon }=
 \sum_{<i,\, j, \, k>} \left(\bf s_{i} \wedge \bf s_{j} \, +
\bf s_{j} \wedge \bf s_{k} \, +
\bf s_{k} \wedge \bf s_{i} \right)
\label{eq-helicity}
\end{equation}
where the sum is taken on upward triangles in the
counter-clockwise direction, the third angle locates the direction
of the magnetization of one sublattice.
In this paper we shall call ($\bf 3$) the direction of the helicity,
($\bf 1$) the direction of the $A$-sublattice magnetization  and ($\bf 2$)
the third orthogonal direction.
When the reference to a laboratory frame will be necessary, we shall add
a prime to the  N\'eel-axes frame indices, keeping the unprimed
quantities for the laboratory frame.
In the N\'eel ground-states (and in their first long-wavelength
excitations) the length of the sublattices magnetization is supposed to
be constant; thus, the orientations of the N\'eel frame are the only
variables of the problem (homogeneous on the lattice in the ground-states
and slowly spatially variable in the first excitations): in other words
the order parameter is an element of $SO(3)$.
Let us now consider the dynamics of this collective
variable in the homogeneous N\'eel state.
On a finite lattice, this collective variable has
a finite ``inertia'' and its free dynamics  is entirely
determined by its angular nature and the
isotropy of spin space. We thus expect it to be that of a free top.
Such a dynamics is at best described in the frame of principal axes of
the object. The planar symmetry of the  N\'eel state implies that one
of the principal axes will be directed perpendicular to the plane of
the spins, that is parallel to  the helicity ${\bf \Upsilon }$.
In the principal axes of the  N\'eel state, the hamiltonian describing
the free dynamics of the system reads:
\begin{equation}
\label{eq-top}
H_{\rm eff}= \frac{S_{1}^{2}}{2I_{1}}
	   + \frac{S_{2}^{2}}{2I_{2}}
           + \frac{S_{3}^{2}}{2I_{3}}
\end{equation}
where $ S_{1}, S_{2}, S_{3}$ are the
three components of the {\sl total}
spin of the system and $ I_{1}, I_{2}, I_{3}$ the principal moments of
inertia.
 $ I_{1}, I_{2}, I_{3}$ are indeed linear responses to homogeneous
magnetic fields in the spin plane ($ I_{1}, I_{2}$) or perpendicular
to it ($ I_{3}$). They are the extensive homogeneous susceptibilities
of this system.
In the thermodynamic limit, we expect $ I_{1}$ and $ I_{2}$ to be
equal by symmetry, as they are associated to magnetic fields which
rotate the spins out of the plane:
we denote them $ I_{\perp} \equiv N \chi_{\perp}$, and $ I_{3}$ is
denoted $ I_{\parallel} \equiv N \chi_{\parallel}$, where $\chi_{\perp}$
and $\chi_{\parallel}$ are the perpendicular and parallel
susceptibilities.
  For the classical Heisenberg
  model,  $ \chi_{\perp} =\chi_{\parallel}=1/18 $ (see Eq.\ref{eq-a-11}).
Nevertheless, it is likely
that on the triangular lattice the renormalization of the two
quantities by quantum fluctuations can be
different\cite{dr89,adm93,css294,css94}.
Thus, we shall focus on the
dynamics of a symmetric top.

All the considerations up to now lay on the angular nature of
the collective variable and are valid in a classical as well
as in a quantum mechanics point of view.
The quantization rules can
be obtained from elementary quantum mechanics, but
it is interesting to look first at the classical
approach of the dynamics of this top
\cite{ll166}. The free
dynamics of a top is entirely described by the relative motion of three
directions (see fig.\ref{fig:top}): the total angular momentum $\bf S $
(a conserved quantity which is fixed in the lab frame), the angular
rotation vector $\bf \Omega $ and the principal axis of inertia of
the top (axis $\bf 3^{'}$, directed along the helicity of the
N\'eel state). For a free symmetric top, these three directions are
in fact always coplanar, and the global motion is the combination of
two rotations: the precession of axis  $\bf 3^{'}$ of the top
around the total angular momentum $\bf S $  with an angular velocity
   $\Omega_{pr} = S/I_{1}$
and the uniform spinning of the top on itself around
 $\bf 3^{'}$ with the angular velocity
   $\Omega_{\bf 3^{'}} = S_{\bf 3^{'}}/I_{3} = S \cos \theta/ I_{3}$.
The dynamics reduces to two separate motions of two angular
variables constrained to vary on $ [0, 4\pi] $  ($ SU(2) $ variables).
In quantum mechanics, these two angular constraints imply
the quantizations of the eigenstates of the associated rotation
generators.
The first condition provides us trivially with the
quantization of the total angular momentum, and its ($ 2S+1 $)
degeneracy.
The second condition is not so trivial and implies the quantization
of the projection of the total spin on the $\bf 3^{'}$ axis of the
top (not to be confused with the projection of the total spin on the
 $\bf 3 $ lab-frame axis). For a given $S$ value, $ S_{3'}$
can thus take $ 2S+1 $ values ranging from $S$ to $ -S $.
The hamiltonian then appears in the canonical form:
\begin{equation}
 H_{\rm eff}= \frac{{\bf S}^{2}}{2I_{\perp}}
	    + S_{ 3'}^{2} (\frac{1}{2I_{\parallel}} -\frac{1}{2I_{\perp}})
\label{eq-heff}
\end{equation}
The degeneracy of the eigenlevels of Eq.\ref{eq-heff}
is ($ 2S+1 $) for $ S_{3'}=0 $ and $ 2 \times (2S+1) $ for
 $ S_{3'}\neq 0 $; for a given $S$ value of the total momentum the
dimension of the Hilbert space is $ (2S+1)^{2}$.

Thus the ${\sl rigidity}$ of the ordered N\'eel states,
implies that the low lying spectrum of Eq.\ref{eq-heis} must
map {\sl in the thermodynamic limit},
on the spectrum of Eq.\ref{eq-heff}: that is indeed a very
striking feature of finite-size spectra displayed in
fig.\ref{fig:transitions} and \ref{fig:quantum top}
where both the leading behavior of
Eq.\ref{eq-heff} and the global multiplicity $ (2S+1)^{2}$ of the top
is well verified up to a total spin $ S=N/6 $ (for
higher values of $S$, the multiplicity $(2S+1)N_S$
($N_S$ given by Eq.\ref{eq-muls}) is lower than $ (2S+1)^2 $).
The inertia $ I_{\perp}$ in Eq.\ref{eq-heff} is indeed
an extensive quantity scaling as $ N $ (see fig.\ref{fig:inertia}).
This leads to a determination of ${\chi}_{\perp} = I_{\perp}/N $ which
is not very different, and only a bit smaller than the first-order
spin-wave calculations of Chubukov et al.\cite{css94}
(see fig.\ref{fig:suceptperp}a).

The situation is more difficult as regard
the precise determination of the anisotropy and of ${\chi}_{\parallel}$.
For the small samples studied in this work, the size effects are still
extremely important:
for $ N=9 $ the top is spherical, for $ N=12 $ no definite sign can be
ascribed to $ (\frac{1}{2I_{\parallel}} -\frac{1}{2I_{\perp}}) $.
For the two larger samples ($N=21,27$) there is a tendency towards
a behavior of an oblate top
 $ (\frac{1}{2I_{\parallel}} -\frac{1}{2I_{\perp}})<0 $,
but the expected degeneracies of the symmetric top are not
present leading to large uncertainties
on  ${\chi}_{\parallel}$ (see fig.\ref{fig:quantum top}) .
Unfortunately the Hilbert space of the quantum top in the $ N=36 $ case
is too large to allow the determination of the complete spectrum
of low lying levels and we are thus unable to decide clearly even
on the sign of the above quantity in this case.

\subsection {Spatial symmetries of the quantum effective top}

We have  determined in subsection III.B the three IRs
characterizing N\'eel order on the triangular lattice:

 $\Gamma_1 $ : [ ${\bf k} = {\bf 0}$, ${\cal R}_\pi\Psi=\Psi $, ${\cal
R}_{2\pi/3}\Psi=\Psi $,
 $\sigma_x\Psi=\Psi $ ],
 $\Gamma_2 $ : [ ${\bf k} = {\bf 0}$, ${\cal R}_\pi\Psi=-\Psi $, ${\cal
R}_{2\pi/3}\Psi=\Psi $,
                                     $\sigma_x\Psi=\Psi $ ]  and
 $\Gamma_3 $ : [ ${\bf k} =  \pm {\bf k}_{0}$, ${\cal R}_{2\pi/3}\Psi=\Psi $,
$\sigma_x\Psi=\Psi $ ].

These IRs should be identified with the three IRs of the invariance
group  $ C_{3v}$ of the magnetic arrangement.
Each symmetry of the lattice results in a permutation of the
magnetic sublattices. Because the order parameter is an element
of $SO(3)$, such a transformation
may be cleared by a global rotational symmetry
of the spins. This property allows us to compute
the number of occurrences of the three IRs (see Appendix B):
\begin{eqnarray}
\label{eq-top-deg}
\!\!\!n_{\Gamma_1}&\!\!\!  =\!\!\! & (a+3b+2c)/6, \; \;
n_{\Gamma_2}=(a-3b+2c)/6, \;
\hbox{ \rm and }\; n_{\Gamma_3}= (a-c)/3, \nonumber \\
 \hbox{ \rm where }\nonumber \\ \!\!\!a & = & 2S+1,\; b=\cos(S\pi) \hbox{ \rm
and }
c=\sin(\frac{2\pi}{3}(2S+1))/\sin(\frac{2\pi}{3}).
\end{eqnarray}
In the hypothesis of an \`a la N\'eel symmetry breaking,
the low energy spectrum of Eq.\ref{eq-heis} should thus contain for
each $S$, $ M_{S}$ subspace:

 $ n_{\Gamma_1}$  [ ${\bf k} = {\bf 0}$, ${\cal R}_\pi\Psi=\Psi $, ${\cal
R}_{2\pi/3}\Psi=\Psi $,
				     $\sigma_x\Psi=\Psi $ ] IR,

 $ n_{\Gamma_2}$  [ ${\bf k} = {\bf 0}$, ${\cal R}_\pi\Psi=-\Psi $, ${\cal
R}_{2\pi/3}\Psi=\Psi $,
				     $\sigma_x\Psi=\Psi $ ] IR and

 $ n_{\Gamma_3}$ couples of degenerate
     [ ${\bf k} =  \pm {\bf k}_{0}$,
     ${\cal R}_{2\pi/3}\Psi=\Psi $, $\sigma_x\Psi=\Psi $ ] IRs.

This appears to be true in all the exact spectra that we have computed,
for all $S$ values up to $ N/6 $
(see Table \ref{table-lowest-energies} and fig.\ref{fig:quantum top}).

The mapping of the low lying levels of Eq.\ref{eq-heis}
on those of Eq.\ref{eq-heff} would imply  a
``quasi-degeneracy'' of some $\Gamma_{1}$ and  $\Gamma_{2}$ levels.
This phenomenon is $\sl not$ $ \sl present $ in the studied
samples (see fig.\ref{fig:quantum top});
this does not exclude the possibility of a quasi-degeneracy
in the thermodynamic limit, but this explains the difficulty to
extract ${\chi}_{\parallel}$ from these data.

At that point, we have determined both the dynamics and the symmetries
of the QDJS that should appear in exact spectra of a system
exhibiting a  N\'eel ordered phase in the thermodynamic limit.
As regards these criteria
the spectra of the THA point in favor of a N\'eel ground-state: the
scaling of ${I}_{\perp}$  seems even to dismiss the case of quantum
criticality (${I}_{\perp}$ should then scale as $ N^{1/2}$ as
shown by Azaria et al\cite{adm93}). We shall come back to this
point in the conclusion.

\subsection {Spectra of samples with $ N= 3p+1$}

The periodic samples which do not frustrate N\'eel order, must have
three sublattices invariant in the periodic boundary conditions, that
is a number of spins $ N=3p $. On the nowadays computers and algorithms,
memory requirements limit the studies to samples of $ N=9,12,21,27 $
and $ 36 $.
In order to enlarge the number of data available and have a complementary check
of our hypothesis we have
thus relaxed the periodic boundary conditions to allow the studies
of the $ 7,13,16,19,25 $ and $ 28 $ samples.
Let us look at such a tiling
of the infinite lattice (see fig.\ref{fig:tiling}).
In order to preserve the possibility
of a three lattices symmetry breaking on the infinite lattice, the
translations operations ${\bf T}_{1}$ and ${\bf T}_{2}$ should
be linked to a rotation of the spins: to be specific, in fig.\ref{fig:tiling},
the interaction
between the spin (10) and (1') is chosen to be equal to the interaction
of (10) with the periodic image of (1) rotated by $ 2\pi/3 $ and so on.
Such boundary conditions, which imply the selection of an axis of rotation,
break the rotational symmetry of the hamiltonian and the total spin
is no longer a good quantum number, but the component $ S_{3}$ of the
total spin on the axis of rotation remains a conserved
quantity. If we keep a global reference frame for the spins,
the translational symmetry is difficult to handle. But this essential
symmetry becomes again trivial by the choice of local frames rotated by
 $\pm 2\pi/3$ (resp. $\mp 2\pi/3$) in one-step translations of
${\bf u}_1$ (resp. ${\bf u}_2$).
Within such a choice the Heisenberg hamiltonian reads
\begin{equation}
{\cal H} = 2 \sum_{<i,j>} {\tilde {\bf s}_{i}}. {\cal R}_{3}^{-1}(2\pi/3)
{\tilde {\bf s}_{j}} {\cal R}_{3}(2\pi/3),
\label{eq-hamiltourn}
\end{equation}
where $\tilde{{\bf s}_{i}}$, $\tilde{{\bf s}_{j}}$ are the spin operators in
the local reference frames and ${\cal R}_{3} (2\pi/3) = \exp (-i\pi/3
\sigma_3)$,  $\sigma_3$ being the third Pauli matrix.
We can in fact solve this problem for any value ($\phi, \psi $)
of the rotations associated to one step translations in the
 ${\bf u}_1$  and ${\bf u}_2$ directions (as long as the
total rotation around a plaquette is zero). We have indeed done it
and verified that the absolute minimum of the energy is obtained
for the values ($\pm 2\pi/3$) for samples with $ N=3p+1 $ and
for 0 and $\pm 2\pi/3 $ for samples with $ N=3p $
(see fig.\ref{fig:twist-ener}).
 The same method
gives information on the spin stiffnesses of the system\cite{lblp94}.

If the system has N\'eel order at $T = 0 $, the spectrum of the
low lying levels of Eq.\ref{eq-hamiltourn} is easily deduced from
the general considerations of section III.C. The present boundary
conditions induce a cylindrical symmetry of the problem around
the $\bf{3}$ axis of the laboratory: this implies that the helicity axis
 $\bf{3^{'}}$ of
the N\'eel state (see fig.\ref{fig:top}) is now constrained to coincide
with the $\bf{3}$ axis of the laboratory. There is only one
degree of freedom left, associated with the rotations around this axis,
and one constant of motion: the $S_3$ component of the total
spin. The effective  hamiltonian Eq.\ref{eq-top} in these
conditions reduces to:
\begin{equation}
{\cal H}_{\rm eff} = \frac{S_{3}^{2}}{2I_{3}} =
\frac{S_{3}^{2}}{2N \chi_{\parallel}}
\label{eq-hamileff1}
\end{equation}
the two other terms in Eq.\ref{eq-top} not related to a
constant of motion should average to zero in any eigenstates.
In our problem, $S_3$ can take all the values ranging from
 $-N/2$ up to $N/2$ and the degeneracy of each eigenstates of
Eq.\ref{eq-hamileff1} is 2 for $S_3 \ne 0$ and 1 for
$S_3 = 0$.
Diagonalization results entirely
corroborate these deductions as can be seen in fig.\ref{fig:en-nonmult}.
 The
analysis of the low lying levels obeying Eq.\ref{eq-hamileff1}
gives a determination of $\chi_{\parallel}$ (see fig.\ref{fig:suceptperp}b).
As for the $N=3p$ case, the
values obtained by this method are in agreement and only
a bit larger than the first-order spin-wave approximation,  moreover the
size effects seem to be roughly the same in the two approaches.

\subsection {Extrapolation of the ground-state energy per bond}

For ordered systems, the finite-size effect on the
$N$-ground-state is mainly due to
the cut-off of the long wave-length excitations. In the supposed-to-be
N\'eel order these excitations are magnons, linear in $k$ and
the leading term of the finite-size corrections to the ground-state energy
is of order $ N^{-1/2}$ and thus the ground-state energy
per bond varies as $ N^{-3/2}$ : more sophisticated approaches would
allow to compute the first coefficients of these expansions as function of
the spin-wave velocities\cite{nz89,hn93,adm93}. In fact in the previously
developed theories\cite{dr89,adm93} only the $ N=3p $ samples
have been taken into consideration: in this case the fluctuations
associated to the three modes of magnons equally contribute to the
renormalization. In the $ N=3p+1 $ samples, the approach to the
singular points at ${\bf k} = {\bf 0}$ and $\pm {\bf k}_{0}$ is
different and we expect different coefficients in the $ N^{-\alpha}$
expansion. It is effectively what is seen in the finite-size
scaling analysis of the spin-wave susceptibilities (fig.\ref{fig:suceptperp})
and of the spin-wave energies (fig.\ref{fig:finit-en}).

The analysis of the diagonalization results is more
subtle, because it depends on the parity of $N$. The ground-state
is either an $ S=0 $ or an $ S=1/2 $ state and in this latter case one
has to take into account the top inertial effects described in the
previous subsections. This leads to an $S/N$
correction to the ground-state energy (an
 $ S/N^2$ correction to the energy per bond) which is noticeable
for the small sizes considered here. In order to compare the
exact results to finite-size spin-wave results, it is
necessary to extrapolate the exact $ E(S,N) $ to an effective value
 $ E(0,N) $. Two estimates of $ E(0,N) $ can be computed. The first one
is obtained by the subtraction from the ground-state energy
 of the main inertial contribution of $ < \Psi_{0} | {\cal H}_{\rm eff}
| \Psi_{0} > $ that is for $ N=3p $ samples:
\begin{equation}
< 2 {\bf s}_{i}.{\bf s}_{j} >_{0,N}^{[1]} = < \Psi_{0} |
2 {\bf s}_{i}.{\bf s}_{j} | \Psi_{0} > - < \Psi_{0} |
\frac{{\bf S}^2}{6N^2 \chi_{\perp}} | \Psi_{0} >
\end{equation}
and for  $ N=3p +1 $ samples
\begin{equation}
< 2 {\bf s}_{i}.{\bf s}_{j} >_{0,N}^{[1]} = < \Psi_{0} |
2 {\bf s}_{i}.{\bf s}_{j} | \Psi_{0} > - < \Psi_{0} |
\frac{S_{3}^2}{6N^2 \chi_{\parallel}} | \Psi_{0} >.
\end{equation}
The second estimate is an average of the same quantity taken
on all the QDJS up to $ S=N/6 $
\begin{equation}
< 2 {\bf s}_{i}.{\bf s}_{j} >_{0,N}^{[2]} =
1/N_{QDJS} \sum_{S=min}^{S=N/6} < QDJS | \frac{{\cal H} - {\cal H}_{\rm eff}}
{3N} | QDJS >
\end{equation}
where ${\cal H}$ is the Heisenberg hamiltonian Eq.\ref{eq-heis}
and ${\cal H}_{\rm eff}$ is given by Eq.12 or Eq.15 according to the
number of spins in the samples.

The first estimate (used in \cite{blp92}) is different from the
exact ground-state for odd N samples only, whereas the second estimate
which is an average over a large number of levels, always differ
from the exact ground-state. These averaged results are compared
to finite-size spin-wave results (see fig.\ref{fig:finit-en}) where it is seen
that the exact results are lower than the spin-wave results,
the finite-size effects on the two sets of data being nevertheless roughly
the same. It has been shown  by Azaria and coworkers that the
finite-size correction to the ground-state energy of the $ N=3p $ samples
is\cite{adm93}:
\begin{equation}
E_{N} = E_{\infty} -\alpha (c_{\parallel}+2c_{\perp}) N^{-3/2}
\label{eq-etrap}
\end{equation}
where $\alpha $ is a geometrical form factor and
 $ c_{\parallel}, c_{\perp}$ are in plane and out of plane
spin-wave velocities.

On the basis of our numerical data we cannot decide on the
exact value of the renormalized spin-wave velocities: it does
not seem to be largely different from the spin-wave results, both in the $ N=3p
$ (fig.\ref{fig:finit-en}a) and in
the $ N=3p+1 $ cases (fig.\ref{fig:finit-en}b).
The complete set of data analyzed within the following hypothesis
\begin{equation}
c_{\parallel} = c_{\parallel sw} \;\; \;\;
c_{\perp} = c_{\perp sw}
\end{equation}
gives an estimate of the energy  per bond in the thermodynamic
limit
\begin{equation}
< 2 {\bf s}_{i}.{\bf s}_{j}>_{\infty} = -0.363
\end{equation}
(see fig.\ref{fig:finit-en}c).

At the end of this section, we can conclude that the spectral
properties of the Heisenberg hamiltonian give strong argument
in favor of an ordered ground-state and that the numerical
spectral data and finite-size corrections are consistent with
the picture and not very different from the spin-wave results.
We will now discuss the important issue of the value of the
order parameter of these N\'eel states.

\section{ Order parameter of the N\'eel states}

The check for long range order on the lattice could involve
the measurement of the two-points correlations
$< {\bf s}_{i}(0).{\bf s}_{j}(r)>$
or that of a macroscopic observable as the sublattice magnetization.

In view of the lattice sizes that can be studied, the analysis of
 the two-points correlation functions
is rather unconclusive, the asymptotic behavior is far from reached
and the sizes are still too small to check the Kennedy-Lieb-Shastry
inequality\cite{kls88}.
The measurement of the squared sublattice magnetization ${\cal M}^{2}$:
\begin{equation}
\label{eq-subl-magn}
{\cal M}^{2} = < {\bf S}^{2}_{A}> = < {\bf S}^{2}_{B}> = < {\bf S}^{2}_{C}>
\end{equation}
or of the static structure factor ${\cal S}( {\bf k}) $ defined in
state $ |\Psi> $ as
\begin{equation}
      {\cal S}( {\bf k})  =   <\Psi|{\bf  S}_{-k} .{ \bf S}_{k}|\Psi >
      \label{eq-struct-fac}
   \end{equation}
(with ${\bf  S}_{k}$ in Eq.6)
is easier. ${\cal S}( {\bf k}) $ measured in the ground-state
of the $N=3p$ samples is shown in fig.\ref{fig:static-fact}.
The signature of N\'eel order in the $N$-ground-state should
appear as a macroscopic value of ${\cal M}$ or
${\cal S}( {\bf k_{o}})$ using the following identity:
\begin{equation}
 2\,\,N \,\,{\cal S}({\bf k}_{0}) + <\Psi|{\bf S}^{2}|\Psi > = 9\, {\cal M}^{2}
\end{equation}

It should be kept in mind that,
in the studied samples, the largest sublattices spins
 range from 3/2 (for $N=9$) to 6
(for $N=36$): they cannot be treated
as classical quantities. ${\cal M}^{2}$
 is the square of a vector, it should be normalized by its maximum
quantum value $(N/6)(N/6+1)$
(reached in the fully aligned classical N\'eel state
$|\Psi_0^0(i, S, M_{S})>$ of subsection III.A).
The structure factor (same tensorial
form as ${\cal M}^{2}$) should be identically normalized.
Two definitions of the order parameter ($X$ or $Y$) seem equally valid:
\begin{equation}
 \label{eq-order}
X^2 = \frac {36{\cal M}^{2}}{ N(N+6)},  ~~~~~~~{\rm or} ~~~~~~~
Y^2 =8\frac {{\cal S}( {\bf k}_{0})}{N+6},
 \end{equation}
 $X$ saturates at one in the classical N\'eel state (and in the
 $N=9$ sample where the quantum fluctuations are ineffective to
reduce the magnetization sublattices)
and should be decreased by the quantum fluctuations in the
quantum ground-state
\footnote{Examination of this problem of normalization
has been underestimated in previous studies\cite{nn88,lr93}.
Because of an erroneous normalization most of the
weight of the extrapolation procedure has been carried on unphysical numbers
larger than the saturation value of the fully aligned classical
N\'eel state. This essentially explains the difference in the
conclusions of the previous authors and ourselves as regard the
question of order of the THA. Deutscher and Everts have been
faced with the same kind of difficulty in their recent work on the
 $ J_{1}-J_{2}$ hamiltonian on the triangular lattice\cite{de93}; by studying
the behavior of the system when $ J_{2}/J_{1} \rightarrow -\infty $
they arrive at the same kind of renormalization to order $ 1/N $
than ours. The above line of reasoning is probably more general.}.
For the $ N=3p+1 $ samples, it is impossible to define a
sublattice magnetization. On the other hand one can measure the
sum $\vec{{\cal M}_{t}}$ of the magnetizations computed in the local basis:
for a fully aligned N\'eel state the square of this vector should
saturate to $ (N/2)(N/2 + 1) $. We thus define in this case:
\begin{equation}
 \label{eq-order1}
X^2 = \frac {4{\cal M}_{t}^{2}}{ N(N+2)}
 \end{equation}

Finite-size scaling analysis indicates that the leading correction
to these parameters should go as $ N^{-1/2}$; analysis of the results
(see fig.\ref{fig:QDJ-OP})
shows that for these small values of $ N $ the subleading
correction is important: the $ N \rightarrow \infty $ extrapolation
is thus rather difficult, but an extrapolation to a zero value
seems highly improbable.

On the other hand, the comparison between the diagonalization results
and first-order spin-wave results is interesting: it is seen in
fig.\ref{fig:QDJ-OP}
that the two sets of results do not differ by large amounts.
It should be noted that Miyake spin-wave results are
in our point view incorrect for finite sizes\cite{m92}. We agree with
Deutscher and Everts formula for first-order spin-wave results which
restrict the renormalization by quantum fluctuations to
 ${\bf k} \ne 0,\pm {\bf k}_{0}$ wave vectors\cite{de93}. As it is well
known the spin-wave hamiltonian cannot be bosonized for ${\bf k}
= 0,\pm {\bf k}_{0}$ (the Bogolioubov transformation becomes
singular). It is a bit lengthy but straightforward to show that
these three Fourier components of the spin-wave hamiltonian can
be recombined to give the total spin $S^2$ and
that they do not participate to the renormalization of the order
parameter (such a remark has recently been developed by
Zhong and Sorella in the case of the square lattice Heisenberg
hamiltonian\cite{zs93}). Careful examination of these singular
terms shows that the first-order correction to the ground-state
energy is exactly obtained by the summation of the usual
formula on the $ N $ points of the Brillouin zone:
\begin{equation}
< 2 {\bf s}_{i}.{\bf s}_{j}>_{sw} = -1/4 + \frac{1}{2 N}\sum_{\bf{k}}
( \omega_{\bf k} - 1)
   \label{eq-SWE0}
\end{equation}
where $\omega_{\bf k} = (1-\gamma_{\bf k})^{1/2}(1+2\gamma_{{\bf k}})^{1/2}$
and $\gamma_{\bf k} =  f({\bf k})/6 $  (see Eq.\ref{eq-gam}).
On the other hand, the correction to the ${\cal M}_{z}$ magnetization
involves exclusively the ${\bf k} \ne 0,\pm {\bf k}_{0}$
points of the Brillouin zone that is $N-3$ points in the $N=3p$
samples and $N-1$ points in the $N=3p+1$ samples.
\begin{equation}
< {\cal M}_{z} > = \frac{1}{2} [ 1 -\frac{1}{N} \sum_{\bf{k} \ne
0,\pm {\bf k}_{0}} ( \frac{1+\frac{\gamma_{\bf{k}}}{2}}{\omega_{\bf k}}
-1) ]
   \label{eq-SWOP}
\end{equation}

The careful comparison between the exact
results
and the first-order spin-wave results leads us to conclude
that the spin-wave approximation seems an extremely good
quantitative approximation for the considered sizes;
on the basis of the present data, it seems highly hazardous
to adopt other estimate of the thermodynamic limit
that the spin-wave results\cite{jlg89}
\footnote{Definitions and comparison of order parameter
in symmetry breaking solutions (spin-wave approximation)
and symmetry non breaking one (exact diagonalizations) has
given birth to many studies\cite{khl89,kt93,g66,dls78,w66}.
In this work, we take care to compare two parameters that are
normalized to 1 in the absence of fluctuations and should be
equal in the thermodynamic limit. The problem of the symmetry
breaking will be studied in a forthcoming
paper.}. In a previous paper\cite{blp92},we have shown
that, for a given size, the order parameter modulus is roughly
the same in all the QDJS.

\section{Conclusion}

The thorough analysis of exact results of diagonalization
of the Heisenberg hamiltonian on periodic samples with triangular
symmetry has brought various pieces of information both qualitative
and quantitative.

The qualitative information emerge from the symmetry analysis
of the spectra: this analysis shows a strict
hierarchy of levels. The first family, degenerated to the absolute
ground-state in the thermodynamic limit, contains all the
quantum states and only those needed to build symmetry breaking
N\'eel states with three sublattices magnetizations in a
 $ C_{3v}$ invariant configuration.
Their dynamics can roughly be mapped on the dynamics of
a quantum symmetric top as expected from general considerations
on the nature of the order parameter in the N\'eel ordered states
\footnote{We should mention that the first exhibition of such
a structure of low lying levels invoked by Anderson in his
seminal paper of 1952 appears in a paper by Gross, Sanchez-Velasco
and Siggia\cite{gsvs289} that regrettedly escaped our
vigilance in our preliminary report of 1992, indeed most of
the theoretical hints were present in the first paper of
this series.}.
Understanding this family of levels from the coupling of
three macroscopic spins explains why the multiplicity of this
quasi-degenerate set of levels (the QDJS) is of order $ N^{3}$ (in the
square lattice case the multiplicity should be of order
 $ N^{2}$) which gives in the thermodynamic limit an entropy
of order $ \log N $. Rather unexpectedly the present analysis
does not give  quantitative answers extremely different
from first-order spin-wave results. On the basis of the
coherence of the data concerning the symmetries of the QDJS,
their dynamics, the energies, the spin
susceptibilities, and order parameter of both the $ N=3p $ and
 $ N=3p+1 $ samples and  their size effects, one can only
sustain the hypothesis of an ``ordered ground-state'' for these
small samples. Indeed we have no information of the effect
of softer quantum fluctuations with wavelengths larger than
about 6 lattice sizes. The clusters expansions of Singh and Huse
and the high temperature expansions of Elstner, Singh and Young
point to a weaker order (if any?) than obtained in the
first-order spin-wave calculations. In these two last methods
the invoked clusters are smaller than our largest sizes but
there are no bias due to periodic boundary conditions and
the sampling of the quantum fluctuations is different from
what is done in this work.

The present state of the art does not exclude that for
larger sizes than those explored today quantum fluctuations
drive the system towards criticality but there is in the small
samples studied here no evidence of such a behavior.

{\bf Acknowledgements}:

We acknowledge many fruitful discussions with P. Azaria, B. Delamotte,
D. Mouhanna, S. Sachdev, P. Sindzingre and A.P. Young.
We have benefited from a grant of computer time at Centre de Calcul
Vectoriel pour la Recherche (CCVR), Palaiseau, France.

\section{Appendix A: Group Theory on the THA,
numerical method}

Here, we explain how to obtain the complete spectrum of the spin-$ 1/2 $
THA, for the largest possible samples compatible with the
triangular symmetry of the infinite lattice.
   The number of eigenstates increases as $2^N$ with $N$.
   Even for very small values of $N$, there is no stable numerical
   method to calculate directly all the eigenvalues of ${\cal H}$.
   The problems to be handled are memory space requirement, computer
   time, and degeneracy of states.
   These problems can be reduced by an intensive use of space and spin
   symmetries.
   Indeed, when all symmetries are accounted for, one can work in subspaces
   where eigenvalues are no more degenerate and better control on round
   off errors is obtained.
   These subspaces come from the decomposition of the Hilbert space
   according to the irreducible representations (IRs) of the symmetry
   group of the problem.

This symmetry group is the direct product of the space
   symmetry group of the lattice times the spin rotation group $SU(2)$.
   Thus, its IRs are the tensor products of space-group IRs times
   $SU(2)$ IRs.
   We first consider in details the space symmetry, then quickly the
   spin symmetries and explain how they are implemented in the numerical
   approach.

{}~\\
{\bf{The space symmetry group}}

The spins stand on a triangular lattice defined by the basis vectors
   ${\bf u}_{1}$ and ${\bf u}_{2}$ (see fig.\ref{fig:tiling}).
   On the infinite triangular lattice, the group ${\cal G}_\infty $ of
   symmetries, which leave the lattice sites globally invariant and keep
   the vicinity relations, is the semi-direct product of the translation
   group ${\cal T}_\infty $ times the point group ${\cal P}$.
   The translation group ${\cal T}_\infty $ has two generators: the
   translations of vector ${\bf u}_{1}$ and ${\bf u}_{2}$.
   The point group ${\cal P}$ consists of the symmetry transformations keeping
a
   site invariant: its generators are the planar rotations ${\cal R}_\pi $ and
   ${\cal R}_{2\pi/3}$ and the axial reflection $\sigma_x $ with respect to
   ${\bf u}_{1}$ (see fig.\ref{fig:tiling}).
   The point group ${\cal P}$ is isomorphic to ${\cal C}_{6v}$ (or
   ${\cal D}_6$) and has twelve elements: six rotations and six
   reflections.

In order to preserve the translation invariance and to reduce the
   number of degrees of freedom,
   periodic conditions are used, defined by the vectors
   ${\bf T}_{1}=l{\bf u}_{1}+m{\bf u}_{2}$
   and ${\bf T}_{2}=l'{\bf u}_{1}+m'{\bf u}_{2}$ (see fig.\ref{fig:tiling}).
   ${\cal R}_\pi $ is always a symmetry transformation of such a system.
   For ${\cal R}_{2\pi/3}$ to be a symmetry transformation of the
   periodic lattice, we chose
   ${\bf T}_{2}={\cal R}_{\pi/3}{\bf T}_{1}$ ($ l'=-m $; $ m'=l+m$).
   The number $ N $ of sites per cell (${\bf T}_{1},{\bf T}_{2}$)
   is thus given by:
   \begin{equation}
      \label{eq-N}
      N=l^2+lm+m^2.
   \end{equation}
   The system has an axial reflection when $ lm=0 $ or $ l=m $.
   Therefore the point group is ${\cal P}\equiv{\cal C}_{6v}$ (resp.
   ${\cal C}_6 $ if there is no $\sigma_x $ axis) and the space group
   ${\cal G}_N={\cal T}_N\wedge {\cal C}_{6v}$ (resp. ${\cal T}_N\wedge
   {\cal C}_6$) has $ 12N $ (resp. $ 6N$) elements.
   In the following, periodic samples with $ N $ multiple of 3 are used.

Let $ E_N $ be the space of wavefunctions for a system of $ N $ spins.
  A wavefunction is a linear combination of configurations:
  \begin{equation}
      \label{eq-wavef}
      \Psi= \sum_c \alpha_c |c>
  \end{equation}
  where a configuration $ |c> $ is an element of the basis
  ${\cal B}_N=\left\{|\uparrow>,|\downarrow>\right\}^N $ of $ E_N $.
  $ E_N $ is a linear representation of ${\cal G}_N $.
  It can be expressed as a direct sum of IRs of ${\cal G}_N $.
  Summing the isomorphic IRs leads to a unique factorization in a direct
  sum of subspaces $ E_{N,\rho}$ associated to the various IRs $\rho $ of
  ${\cal G}_N $.
  Each of these subspaces is invariant under ${\cal H}$.
  Because ${\cal G}_N $ is not commutative, some of the IRs have dimensions
  greater than one resulting in a degeneracy of ${\cal H}$ on $ E_{N,\rho}$.
  Thanks to the rather simple structure of group ${\cal G}_N $,
  one can eliminate this degeneracy by working on some subspace $ E'_{N,\rho}$
  associated to some one-dimensional IR $\rho $ of some
  subgroup ${\cal G}_{N,\rho}$ of ${\cal G}_{N}$.

By using $ E'_{N,\rho}$ instead of $ E_{N,\rho}$, the eigenvalue
  multiplicities are divided by the dimension of $\rho $, which is also
  equal to the ratio of the cardinal of ${\cal G}_N $ to the cardinal of
  ${\cal G}_{N,\rho}$.

Let us first consider the case where there is no axial reflection:
  the point group ${\cal P}$ is an abelian cyclic group of six elements.
  Because ${\cal T}_N $ is abelian, its IRs are one-dimensional and are
  characterized by ${\bf k}$, a vector of the Brillouin zone.
  Thus, $ E_N $ is factorized in $ N $ subspaces $ E_{N,{\bf k}}$.
  The group ${\cal P}$ acts on these vectors to form different stars.
  Two vectors ${\bf k}_1 $ and ${\bf k}_2 $ belong to the same star if
  it exists an element of ${\cal P}$ (here a rotation) which transforms
  ${\bf k}_1 $ in ${\bf k}_2 $.
  Because this transformation commutes with the hamiltonian, the
  eigenvalue spectra of $ E_{N,{\bf k}_1}$ and $ E_{N,{\bf k}_2}$ will be
  identical.
  For a fixed ${\bf k}$, let ${\cal P}_{\bf k}$ be the subgroup of
  ${\cal P}$ which keeps ${\bf k}$ invariant.
  This subgroup is abelian and all its IRs are one-dimensional.
  The cardinal of ${\cal P}_{\bf k}$ is 6 for ${\bf k}=0 $.
  It is 3 for ${\bf k}=\pm{\bf k}_0 $ ($\pm {\bf k}_0 $ are the
  corners of the Brillouin zone); this occurs only when $ N $ is multiple of 3.
  It is 2 if ${\bf k}=-{\bf k}\ne 0 $ (the middle of the side of the Brillouin
  zone) and this occurs only when $ N $ is multiple of 4.
  It is 1 for the other ${\bf k}$ -vectors.
  The different IRs are listed in Table \ref{table-IRSIX}.

When there is an axial reflection ($ lm=0 $ or $ l=m $), the group
   ${\cal P}$ is generated by $\sigma_x $ and ${\cal R}_{\pi/3}$.
   If the group ${\cal P}_{\bf k}$ is abelian, the previous
   construction is applied.
   If the group ${\cal P}_{\bf k}$ is not abelian, some of its IRs are
   two-dimensional.
   This is the case for ${\bf k}=0 $ (${\cal P}_{\bf k}={\cal P}={\cal
   D}_6 $) and ${\bf k}={\bf k}_0 $ (${\cal P}_{\bf k}={\cal D}_3 $).
   The abelian normal subgroups are respectively ${\cal C}_6 $ and ${\cal
   C}_3 $.
   The different IRs are listed in Table \ref{table-IRDOUZE}.

For small systems, other accidental space symmetries can occur which
   imply new degeneracies.
   If they commute with all the previous space symmetries, the
   degenerate states appear in the same IR.
   Otherwhise, the degenerate states stand in different IRs.

{}~\\
{\bf{The spin symmetry group}}

The hamiltonian of Eq.\ref{eq-heis} is invariant under global spin
   rotations: it commutes with ${\bf S}^2 $, where ${\bf S}$ is the
   total spin.
   For spin-1/2, the symmetry group is $ SU(2) $.
   The IRs are labeled by $S$.

{}~\\
{\bf{Numerical method}}

A general wavefunction is defined by the knowledge of the set of $ 2^N $
   coefficients $\{\alpha_c\}$ Eq.\ref{eq-wavef}.
   The first easy reduction of this set is to work in an eigenspace of
   $ S_z $ : the basis size thus becomes $ C_N^{N/2-S_z}$.
   The next step is to work in a given $ E'_{n,\rho}$, characterized by
   the phase factors associated to each element of ${\cal G}_{N,\rho}$.
   One groups together the configurations in conjugate classes:
   $ |c> $ and $ |c'> $ belong to the same class if it exists an element
   $ X $ of ${\cal G}_{N,\rho}$ such that $ |c'>=X|c> $.
   So, $\alpha_{c'}$ differs from $\alpha_c $ by a
   phase factor listed in Tables \ref{table-IRSIX} or \ref{table-IRDOUZE}.
   The number of independent coefficients is therefore reduced by
   roughly a factor $ N $ (translations) and at maximum by a factor $ 12N $
   (cardinal of ${\cal G}_{N}$).
   The memory requirement and computer time are reduced by the same
   amount.

Implementation of the $ SU(2) $ symmetry is done via a projector technique
   by using the operator:
   \begin{equation}
       \prod_{\stackrel{i=N/2,N/2-1,\ldots}{i\ne S}}
          \frac{{\bf S}^2-i(i+1)}{S(S+1)-i(i+1)}
   \end{equation}
   where ${\bf S}^2 $ is computed thanks to relation
   ${\bf S}^2=S_z^2+S_z+2S_-S_+ $ and
   $ S_+ $ and $ S_- $ are computed like ${\cal H}$.
   Applying ${\bf S}^2 $ is therefore as fast as applying the
   hamiltonian, or so.

Lancz\"os method has been applied to diagonalize the hamiltonian in the
   subspace associated to each IR $\rho $.
   Working with classes instead of configurations insures that the
   vectors generated by Lancz\"os method stay in the same subspace.
   For small systems, the hamiltonian and the two operators $ S_+ $ and
   $ S_- $ are tabulated; for large
   systems ($ N=36 $), they have to be computed at each step,
   but vector components are stored in an order allowing
   application of these three operators to be vectorized.
   In any case, diagonal part of the hamiltonian is computed separately
   at once and stored.
   Additional details on the technical tricks will be given elsewhere.

The algorithm is as follows:
   $ i) $ first, for some IR, build the conjugate-class table and
        the phase-factor table;
   $ ii) $ tabulate ${\cal H}$, $ S_+ $ and $ S_- $;
   $ ii) $ choose a random initial vector and project it out in the
   desired subspace of $ S^2 $;
   $ iv) $ apply the Lancz\"os algorithm;
   $ v) $ calculate the eigenvectors, correlations...

In fact, round-off errors propagate very quickly in this algorithm.
   In order to eliminate these errors, computed vectors are
   orthogonalized to all previous ones and projected out in the
   $S$-subspace.
   One can compute the dimension $ m_\rho(S) $ of this subspace and
   verify that, after exactly $ m_\rho(S) $ iterations, the modulus of the
   last vector is zero.
   This is a strong test of this algorithm.

 \section{ Appendix B}

Here, we compute the number of replica of each irreducible
representation of the QDJ states if a N\'eel state occurs.
The idea is to use some symmetries that keep
invariant the classical N\'eel state.
These symmetries are the
compositions of the
permutations of the sublattices times specific spin rotations.

Thus, let us start with an {\sl a priori} classical N\'eel state and its
symmetry group. In our case, the N\'eel state is made of three
sublattices as described in section II.
Let us define the plan
 $ (xy) $ of the classical N\'eel state and $ z $ a perpendicular axis.
The group which permutes the
three sublattices is $ S_3 $ (isomorphic to the dihedral group $ D_3 $).
After a permutation which exchanges the spins between the sublattices
(giving an other N\'eel state), the initial N\'eel
state is recovered if the spins are rotated simultaneously.
For example,
the permutation of two sublattices (say $A$ and $B$) times
a spin rotation of
axis parallel to the spins of the last sublattice (thus $C$)
recovers the initial N\'eel state.
Likewise, a cyclic permutation of the three sublattices times a spin
rotation of $2\pi/3$ along the $z$-axis also recovers the initial N\'eel
state.
Thus, two (resp. three) sublattices permutations are associated to
an half (resp. a third of a) turn around
an axis in the $(xy)$ plane (resp. around the $z$-axis).

In the quantum case, the permutation of the sublattice is one of the
symmetry of the lattice (say,
central symmetry for the two sublattice permutation
and translation of one step for three sublattice permutation)
times a spin rotation in $SU(2)$.
The group of these spin rotations is the dihedral group $D_3$
for integer spins ($SO(3)$),
and is a group of 12 elements for half-integer spins ($SU(2)$),
whose character table is given in Table \ref{table-C12}.
Note that the type of spin rotation associated to each class is specific
of the N\'eel state and is necessary (here also sufficient) to calculate
the number of replicas $n_{\Gamma_i}$ of each irreducible representation
(IR) $\Gamma_i$.
As all the above transformations conserve the total spin $S$, one can thus
calculate the traces of these rotations in the subspace of fixed $S$, $ M_{S}$:
\begin{equation}
n_{\Gamma_i}^{(S)}=\frac{1}{6} \sum_k Tr^{(S)}({\cal R}_k) \chi_i(k) N_{el}(k)
\end{equation}
For integer (resp. half-integer) spins, only the $n_{\Gamma_i}$
(resp. $n_{\Gamma_i'}$) are non zero.

Now, we identify the $\Gamma_i$ IRs with the IRs of the hamiltonian.
Because the QDJS have already been found to belong to three precise
IRs (see section III.B),
this identification is straightforward: $\Gamma_0$ is the trivial
IR, $\Gamma_1$ is the IR odd with respect to the inversion
(${\bf k} = {\bf 0}$,
invariant under a $2\pi/3 $ rotation), and $\Gamma_3$ is the last
doubly degenerate representation
(${\bf k} = {\bf k}_{0}$, invariant under a $2\pi/3$ rotation).
The identification of the $\Gamma_i'$ is the same for half-integer spins
and leads to formulas (13) both for integer and half-integer spins.

\newpage

\newpage


\begin{table}
 \begin{center}
  \begin{tabular}{|r|rlrllrrr|}
   \hline
   $ N $ & $ <2\vec {\bf s}_i.\vec {\bf s}_j> $
   & $S$ & deg. & \multicolumn{2}{c}{$ {\bf k} $} & ${\cal R}_{2\pi/3} $
   & ${\cal R}_{\pi} $ & $\sigma $\\
   \hline
%
9& *     -0.3888889 &  0.5 &  4 &  3 & -3 &  1 &  0 &  1 \\
&       -0.2777778 &  0.5 & 12 &  0 &  3 &  0 &  0 & -1 \\
&       -0.2777778 &  0.5 & 12 &  0 &  3 &  0 &  0 &  1 \\
& *     -0.2777778 &  1.5 &  4 &  0 &  0 &  1 &  1 &  1 \\
& *     -0.2777778 &  1.5 &  4 &  0 &  0 &  1 & -1 &  1 \\
& *     -0.2777778 &  1.5 &  8 &  3 & -3 &  1 &  0 &  1 \\
&       -0.1666667 &  0.5 &  4 &  0 &  0 & -1 &  1 &  0 \\
&       -0.1666667 &  0.5 &  8 &  3 & -3 & -1 &  0 &  0 \\
&       -0.1666667 &  0.5 & 12 &  0 &  3 &  0 &  0 &  1 \\
&       -0.1666667 &  1.5 & 24 &  0 &  3 &  0 &  0 & -1 \\
&       -0.1666667 &  1.5 & 24 &  0 &  3 &  0 &  0 &  1 \\
&       -0.0925926 &  2.5 &  6 &  0 &  0 &  1 &  1 &  1 \\
&       -0.0925926 &  2.5 & 12 &  3 & -3 &  1 &  0 &  1 \\
\hline
%
12& *     -0.4068868 &  0 &  1 &  0 &  0 &  1 &  1 &  1 \\
& *     -0.3589325 &  1 &  6 &  4 & -4 &  1 &  0 &  1 \\
& *     -0.3570639 &  1 &  3 &  0 &  0 &  1 & -1 &  1 \\
&       -0.3538882 &  0 &  2 &  0 &  0 & -1 &  1 &  0 \\
&       -0.3382480 &  0 &  3 &  0 &  6 &  0 & -1 &  1 \\
&       -0.3370959 &  0 &  6 &  2 &  4 &  0 &  0 &  1 \\
& *     -0.3280998 &  1 &  9 &  0 &  6 &  0 &  1 &  1 \\
&       -0.3132131 &  1 & 18 &  2 &  4 &  0 &  0 &  1 \\
&       -0.3024839 &  1 &  9 &  0 &  6 &  0 &  1 &  1 \\
&       -0.2913787 &  0 &  1 &  0 &  0 &  1 &  1 & -1 \\
&       -0.2907218 &  0 &  4 &  4 & -4 & -1 &  0 &  0 \\
&       -0.2907008 &  1 & 18 &  2 &  4 &  0 &  0 &  1 \\
&       -0.2906078 &  0 &  1 &  0 &  0 &  1 &  1 &  1 \\
& *     -0.2862917 &  2 &  5 &  0 &  0 &  1 &  1 &  1 \\
&       -0.2827618 &  1 & 12 &  4 & -4 & -1 &  0 &  0 \\
& *     -0.2817881 &  2 & 10 &  4 & -4 &  1 &  0 &  1 \\
& *     -0.2694317 &  2 & 10 &  4 & -4 &  1 &  0 &  1 \\
&       -0.2693163 &  1 &  9 &  0 &  6 &  0 &  1 & -1 \\
&       -0.2585320 &  1 & 18 &  2 &  4 &  0 &  0 &  1 \\
%
\hline
21& *     -0.3739972 &  0.5 &  4 &  7 & -7 &  1 &  0 &  0 \\
& *     -0.3516974 &  1.5 &  4 &  0 &  0 &  1 &  1 &  0 \\
& *     -0.3511989 &  1.5 &  8 &  7 & -7 &  1 &  0 &  0 \\
&       -0.3511315 &  0.5 & 12 &  3 &  9 &  0 &  0 &  0 \\
&       -0.3500572 &  0.5 & 12 &  1 & -4 &  0 &  0 &  0 \\
& *     -0.3499587 &  1.5 &  4 &  0 &  0 &  1 & -1 &  0 \\
&       -0.3435291 &  0.5 & 12 &  2 & -8 &  0 &  0 &  0 \\
&       -0.3402059 &  0.5 & 12 &  3 &  9 &  0 &  0 &  0 \\
&       -0.3392388 &  0.5 & 12 &  2 & -8 &  0 &  0 &  0 \\
&       -0.3390870 &  0.5 & 12 &  1 & -4 &  0 &  0 &  0 \\
&       -0.3320711 &  0.5 &  8 &  7 & -7 & -1 &  0 &  0 \\
&       -0.3296426 &  1.5 & 24 &  3 &  9 &  0 &  0 &  0 \\
&       -0.3292417 &  1.5 & 24 &  2 & -8 &  0 &  0 &  0 \\
&       -0.3289908 &  1.5 & 24 &  1 & -4 &  0 &  0 &  0 \\
&       -0.3287770 &  0.5 &  4 &  7 & -7 &  1 &  0 &  0 \\
  &&&&&&&& \\
%
& *     -0.3172244 &  2.5 & 12 &  7 & -7 &  1 &  0 &  0 \\
& *     -0.3171493 &  2.5 &  6 &  0 &  0 &  1 & -1 &  0 \\
& *     -0.3123477 &  2.5 &  6 &  0 &  0 &  1 &  1 &  0 \\
& *     -0.3119088 &  2.5 & 12 &  7 & -7 &  1 &  0 &  0 \\
& *     -0.2728708 &  3.5 &  8 &  0 &  0 &  1 &  1 &  0 \\
& *     -0.2728677 &  3.5 & 16 &  7 & -7 &  1 &  0 &  0 \\
& *     -0.2626233 &  3.5 &  8 &  0 &  0 &  1 & -1 &  0 \\
& *     -0.2624250 &  3.5 & 16 &  7 & -7 &  1 &  0 &  0 \\
& *     -0.2587919 &  3.5 & 16 &  7 & -7 &  1 &  0 &  0 \\
   \hline
%
27& *     -0.3734808 &  0.5 &  4 &  9 & -9 &  1 &  0 &  1 \\
& *     -0.3607827 &  0.5 & 12 &  6 & 12 &  0 &  0 &  1 \\
& *     -0.3587615 &  1.5 &  4 &  0 &  0 &  1 & -1 &  1 \\
& *     -0.3586044 &  1.5 &  8 &  9 & -9 &  1 &  0 &  1 \\
& *     -0.3580457 &  1.5 &  4 &  0 &  0 &  1 &  1 &  1 \\
& *     -0.3566086 &  0.5 & 12 &  3 &  6 &  0 &  0 &  1 \\
&       -0.3522638 &  0.5 & 12 &  0 &  9 &  0 &  0 &  1 \\
&       -0.3517280 &  0.5 & 12 &  0 &  9 &  0 &  0 & -1 \\
&       -0.3500536 &  0.5 & 12 &  3 & -12 &  0 &  0 &  1 \\
&       -0.3492247 &  0.5 & 12 &  6 & 12 &  0 &  0 &  1 \\
&       -0.3482682 &  0.5 & 12 &  3 & -12 &  0 &  0 &  1 \\
&       -0.3467737 &  0.5 & 12 &  3 &  6 &  0 &  0 &  1 \\
&       -0.3466080 &  0.5 &  4 &  9 & -9 &  1 &  0 &  1 \\
&       -0.3461131 &  0.5 & 12 &  3 &  6 &  0 &  0 & -1 \\
&       -0.3452044 &  0.5 & 12 &  6 & 12 &  0 &  0 &  1 \\
&       -0.3448586 &  0.5 & 12 &  6 & 12 &  0 &  0 & -1 \\
& *     -0.3448234 &  1.5 & 24 &  3 & -12 &  0 &  0 &  1 \\
& *     -0.3439540 &  1.5 & 24 &  3 &  6 &  0 &  0 &  1 \\
& *     -0.3438163 &  1.5 & 24 &  6 & 12 &  0 &  0 &  1 \\
&       -0.3436128 &  0.5 &  2 &  0 &  0 &  1 &  1 &  1 \\
&       -0.3433155 &  0.5 & 12 &  3 & -12 &  0 &  0 &  1 \\
&       -0.3432208 &  0.5 & 12 &  3 & -12 &  0 &  0 & -1 \\
&       -0.3429195 &  0.5 & 12 &  0 &  9 &  0 &  0 &  1 \\
&       -0.3422904 &  0.5 &  2 &  0 &  0 &  1 & -1 &  1 \\
  &&&&&&&& \\
%
& *     -0.3359171 &  2.5 & 12 &  9 & -9 &  1 &  0 &  1 \\
& *     -0.3358955 &  2.5 &  6 &  0 &  0 &  1 &  1 &  1 \\
& *     -0.3337237 &  2.5 &  6 &  0 &  0 &  1 & -1 &  1 \\
& *     -0.3336028 &  2.5 & 12 &  9 & -9 &  1 &  0 &  1 \\
& *     -0.3072962 &  3.5 & 16 &  9 & -9 &  1 &  0 &  1 \\
& *     -0.3072094 &  3.5 &  8 &  0 &  0 &  1 & -1 &  1 \\
& *     -0.3017642 &  3.5 &  8 &  0 &  0 &  1 &  1 &  1 \\
& *     -0.3016915 &  3.5 & 16 &  9 & -9 &  1 &  0 &  1 \\
& *     -0.3000848 &  3.5 & 16 &  9 & -9 &  1 &  0 &  1 \\
& *     -0.2734043 &  4.5 & 20 &  9 & -9 &  1 &  0 &  1 \\
& *     -0.2733890 &  4.5 & 10 &  0 &  0 &  1 &  1 &  1 \\
& *     -0.2632354 &  4.5 & 10 &  0 &  0 &  1 & -1 &  1 \\
& *     -0.2630937 &  4.5 & 20 &  9 & -9 &  1 &  0 &  1 \\
& *     -0.2593492 &  4.5 & 20 &  9 & -9 &  1 &  0 &  1 \\
& *     -0.2582988 &  4.5 & 10 &  0 &  0 &  1 &  1 &  1 \\
& *     -0.2575865 &  4.5 & 10 &  0 &  0 &  1 & -1 &  1 \\
\hline
%
36& *     -0.3735823 &  0 &  1 &  0 &  0 &  1 &  1 &  1 \\
& *     -0.3667362 &  1 &  2 &  0 &  0 &  1 & -1 &  1 \\
& *     -0.3555115 &  2 &  3 &  0 &  0 &  1 &  1 &  1 \\
&       -0.3517758 &  0 &  1 &  0 &  0 &  1 &  1 &  1 \\
&       -0.3485718 &  0 &  1 &  0 &  0 &  1 &  1 &  1 \\
&       -0.3430154 &  1 &  2 &  0 &  0 &  1 & -1 &  1 \\
&       -0.3400599 &  1 &  2 &  0 &  0 &  1 & -1 &  1 \\
& *     -0.3398439 &  3 &  4 &  0 &  0 &  1 & -1 &  1 \\
&       -0.3398327 &  2 &  3 &  0 &  0 &  1 &  1 &  1 \\
  &&&&&&&& \\
%
 & *     -0.3386439 &  3 &  4 &  0 &  0 &  1 &  1 &  1 \\
 & *     -0.3381808 &  3 &  4 &  0 &  0 &  1 & -1 &  1 \\
 & *     -0.3204399 &  4 &  5 &  0 &  0 &  1 &  1 &  1 \\
 & *     -0.3173265 &  4 &  5 &  0 &  0 &  1 & -1 &  1 \\
 & *     -0.3162180 &  4 &  5 &  0 &  0 &  1 &  1 &  1 \\
 & *     -0.2977774 &  5 &  6 &  0 &  0 &  1 & -1 &  1 \\
 & *     -0.2923236 &  5 &  6 &  0 &  0 &  1 &  1 &  1 \\
 & *     -0.2898917 &  5 &  6 &  0 &  0 &  1 & -1 &  1 \\
 & *     -0.2721993 &  6 &  7 &  0 &  0 &  1 &  1 &  1 \\
 & *     -0.2638050 &  6 &  7 &  0 &  0 &  1 & -1 &  1 \\
 & *     -0.2594966 &  6 &  7 &  0 &  0 &  1 &  1 &  1 \\
 & *     -0.2565594 &  6 &  7 &  0 &  0 &  1 & -1 &  1 \\
 & *     -0.2563581 &  6 &  7 &  0 &  0 &  1 &  1 &  1 \\
\hline
\end{tabular}
\end{center}
\caption
{Lowest energies, degeneracy and quantum numbers for
the samples $N=9 $ , 12, 21, 27, 36.
Components of vectors $k$ are in units of $2\pi/N$.
In the 3 last columns, 1 means invariant under the symmetry, 0 means no
symmetry and -1 means a phase factor under the symmetry ($j$ for the
rotation of $2\pi/3$ and -1 for the two others).
Stars stand for the Quasi-Degenerate-Joint-States (QDJS).
}
\label{table-lowest-energies}
\end{table}

\begin{table}
     \begin{tabular}{|c c | r@{=}l c |}
     \hline
     ${\bf k}$ & ${\cal P}_{\bf k}$ & \multicolumn{2}{c}{$\rho$}
                                                     & multiplicity \\
     \hline
     ${\bf k}={\bf 0}$ & ${\cal P}$ & ${\cal R}_{2\pi/3} \psi $ & $\psi $ & 1
\\
                &         & ${\cal R}_\pi \psi $ & $\pm\psi $ &  \\ \cline{3-5}
                &         & ${\cal R}_{2\pi/3} \psi $ & $ j\psi $ & 2 \\
                &         & ${\cal R}_\pi \psi $ & $\pm\psi $ &  \\
     \hline
     ${\bf k}=\pm{\bf k}_0 $ & $ <{\cal R}_{2\pi/3}> $
                & ${\cal R}_{2\pi/3} \psi $ & $\psi $ & 2 \\ \cline{3-5}
                &         & ${\cal R}_{2\pi/3} \psi $ & $ j\psi $ & 4 \\
     \hline
     ${\bf k}=-{\bf k}\ne{\bf 0}$ & $\{Id,{\cal R}_{\pi}\}$
                                   & ${\cal R}_\pi \psi $ & $\pm\psi $ & 3 \\
     \hline
     other ${\bf k}$   & $\{Id\}$    & \multicolumn{2}{c}{  } & 6 \\
     \hline
     \end{tabular}
     \caption[99]{Irreducible Representations (IR)
        when there is no axial reflection
        ($ lm\ne0 $ or $ l\ne m$).
        First column: vector of the Brillouin zone.
        Second column: subgroup which keeps this vector invariant;
               $ <{\cal R}_{2\pi/3}> $ stands for the group generated by
               ${\cal R}_{2\pi/3}$.
        Third column: IR list of this subgroup, with the phase factor
                       associated to the transformation;
                       $\psi $ is a wavefunction.
        Fourth column: multiplicity of each eigenvalue found in this IR.
        }
  \label{table-IRSIX}
\end{table}

\begin{table}
     \begin{tabular}{|c c | r@{=}l c |}
     \hline
     ${\bf k}$ & ${\cal P}_{\bf k}$ & \multicolumn{2}{c}{$\rho$}
                                                    & multiplicity \\
     \hline
     ${\bf k}={\bf 0}$ & ${\cal P}$ & ${\cal R}_{2\pi/3} \psi $ & $\psi $ & 1
\\
                &            & ${\cal R}_\pi \psi $ & $\pm\psi $ &  \\
                &            & $\sigma_x \psi $ & $\pm\psi $ & \\
                \cline{3-5}
                &            & ${\cal R}_{2\pi/3} \psi $ & $ j\psi $ & 2 \\
                &            & ${\cal R}_\pi \psi $ & $\pm\psi $ &  \\
     \hline
     ${\bf k}=\pm{\bf k}_0 $ & $ <{\cal R}_{2\pi/3},\sigma_x> $
                & ${\cal R}_{2\pi/3} \psi $ & $\psi $ & 2 \\
                &            & $\sigma_x \psi $ & $\pm\psi $ & \\
                \cline{3-5}
                &            & ${\cal R}_{2\pi/3} \psi $ & $ j\psi $ & 4 \\
                &            & $\sigma_x{\cal R}_\pi \psi $ & $\pm\psi $ &   \\
     \hline
     ${\bf k}=-{\bf k}\ne {\bf 0}$ & $<Id,{\cal R}_{\pi},\sigma_{\bf k}>$
                & ${\cal R}_\pi \psi $ & $\pm\psi $ & 3 \\
             &  & $\sigma_{\bf k} \psi $ & $\pm\psi $ &   \\
     \hline
     $\sigma_{\bf k}{\bf k}={\bf k}\ne {\bf 0}$ & $\{Id,\sigma_{\bf k}\}$
                & $\sigma_{\bf k} \psi $ & $\pm\psi $ & 6 \\
     \hline
     other ${\bf k}$   & $\{Id\}$    & \multicolumn{2}{c}{ } & 12 \\
     \hline
     \end{tabular}
     \caption[99]{
        Idem as in Table \ref{table-IRSIX}, when there is an axial reflection
        ($ lm=0 $ or $ l=m$).
        }
     \label{table-IRDOUZE}
\end{table}

\begin{table}
\begin{tabular}{|c|r r r r r r |}
\hline
 & $ I $ & $ -I $ & $ p_3 $ & $ -p_3 $ & $ p_2 $ & $ -p_2 $\\
 $ N_{el}$ & 1 & 1 & 2 & 2 & 3 & 3 \\
\hline
 $\Gamma_0 $ & $ 1 $ & $ 1 $ & $ 1 $ & $ 1 $ & $ 1 $ & $ 1 $\\
 $\Gamma_1 $ & $ 1 $ & $ 1 $ & $ 1 $ & $ 1 $ & $ -1 $ & $ -1 $\\
 $\Gamma_2 $ & $ 2 $ & $ 2 $ & $ -1 $ & $ -1 $ & $ 0 $ & $ 0 $\\
 $\Gamma_0' $ & $ 1 $ & $ -1 $ & $ 1 $ & $ -1 $ & $ i $ & $ -i $\\
 $\Gamma_1' $ & $ 1 $ & $ -1 $ & $ 1 $ & $ -1 $ & $ -i $ & $ i $\\
 $\Gamma_2' $ & $ 2 $ & $ -2 $ & $ -1 $ & $ 1 $ & $ 0 $ & $ 0 $\\
\hline
 $\phi $ & $ 0 $ & $ 2\pi $ & $ 4\pi/3 $ & $ 2\pi/3 $ & $\pi $ & $\pi $\\
 $ Tr^{(S)}{\cal R}(\phi) $ & $ d $ & $ (-1)^{2S}d $
	& $\frac{\sin(\frac{2\pi}{3}d)}{\sin(\frac{2\pi}{3})}
        $ & $\frac{\sin(\frac{2\pi}{3}d)}{\sin(\frac{2\pi}{3})}
	 $ & $\sin(\frac{\pi}{2}d) $ & $\sin(\frac{\pi}{2}d) $\\
\hline
\end{tabular}
\caption[99]{Character table for the spin-rotation group of $SU(2)$ involved
in the N\'eel state; $p_2$ and $p_3$ denote the class of the spins rotations
associated to 2 and 3 sublattice permutations.
The number of elements in each class is $ N_{el}$.
The rotation angle associated to each class is $\phi$ and the trace
of the rotation in the subspace of fixed spin $S$ is given in the last line,
where $d=2S+1$.
}
\label{table-C12}
\end{table}

\newpage

\begin{figure}
\caption[99]{The classical N\'eel  ground-state: on each $ijk$ triangle
$ {\bf s_{i}} + {\bf s_{j}} + {\bf s_{k}}= {\bf 0}$. This defines three
sublattices $A,B,C$ on which the spins are ferromagnetically aligned;
the angle between the spins of two sublattices is $ 2 \pi /3$. For a
given planar upwards triangular plaquette described in the
counter-clockwise direction, the spins can rotate clockwise or
counter-clockwise, corresponding to the two different helicities:
here a positive helicity is assumed.}
\label{fig:classical}
\end{figure}

\begin{figure}
\caption[99]{Energy spectra of Eq.\ref{eq-heis} versus
${\bf S}^2=S(S+1)$ ;
$a)$ complete spectrum for $N=9$ and 12;
$b)$ lowest energies for $N=21$ and $N=27$.
The horizontal and vertical scales have been enlarged by the same
factor so that the slope of the energy per bond versus S(S+1)
can be compared. One sees on these graphs that this slope goes
rapidly to zero. The straight line is a guide for the eye to
link the low lying energy levels called QDJS (for quasi-degenerate
joint states).}
\label{fig:spectra}
\end{figure}

\begin{figure}
\caption[99]{Moment of inertia versus sample size}
\label{fig:inertia}
\end{figure}

\begin{figure}
\caption[99]{$a)$ Enlargement of the low energy levels for
$N=9$ versus $S(S+1)$. Three-legs-symbols family and - family are
essential ingredients of an \`a la N\'eel symmetry breaking. They
respectively span the subspace of quasi-classical N\'eel ground-states
and the subspace of the long wave-length ${\bf k}\neq {\bf 0}$ excitations:
i.e. the magnons.
Black triangles represent the chiral states.
Open triangles are states invariant under rotation of $2\pi/3$ and odd under
axial symmetry.
$b)$ same as in $a)$ for 27 spins.
$c)$ same as $b)$ with
$\frac{S(S+1)}{6N I_{\perp}}$
subtracted from the whole spectrum.
}
\label{fig:transitions}
\end{figure}

\begin{figure}
\caption[99]{Schematic of the free dynamics of a symmetric top}
\label{fig:top}
\end{figure}

\begin{figure}
\caption[99]{Comparison between the exact spectrum
and a fitted symmetric top
for the $N=27$ sample
: 1, 2 and 3 stand for the $\Gamma_{1}$, $\Gamma_{2}$,
$\Gamma_{3}$ IRs (see text);
$\Box$ : spectrum of an ideal symmetric top
(the isotropic term $ \frac{S(S+1)}{6N I_{\perp}} $
is subtracted from the spectrum to focus the comparison
on the second term of Eq.12);
For $N=27$, all levels of the ideal top ($ \Box $) are doubly
degenerate (we do not take into account the trivial $2S+1$
magnetic degeneracy). In the exact spectrum, the levels 1 and 2
are simply degenerate and the 3 levels are doubly degenerate.
Perfect agreement between the two spectra would necessitate a
quasi degeneracy of 1 and 2 levels.}
\label{fig:quantum top}
\end{figure}

\begin{figure}
\caption[99]{Spin susceptibilities of the THA on finite
samples.
Fig.7a:
$\chi_{\perp}$ (triangles) and $\chi_{\parallel}$ (square)
normalized by its classical value
for the $N=3p$-samples;
large symbols stand for results obtained from the QDJS analysis;
small symbols show the finite-size results of
spin-wave calculation of Chubukov et al\cite{css94}.
Points
represent spin-wave results and indicate the infinite size extrapolation.
Fig.7b: $\chi_{\parallel}$ of the $N=3p+1$-samples; symbols are
the same as in $a)$.}
\label{fig:suceptperp}
\end{figure}

\begin{figure}
\caption[99]{The $N=13$ sample and the tiling of the infinite
lattice. Numbers stand for the spin. Prime (resp. double prime) means
that the original spin is rotated by $\pm  2 \pi/3$ (resp. $\pm  4 \pi/3$).}
\label{fig:tiling}
\end{figure}

\begin{figure}
\caption[99]{Opposite of the ground-state energy of the $N=21$
and $N=19$ samples as a function of the spin-rotation angles
$\phi$, $\psi$ attached to the translations ${\bf u}_1$, ${\bf u}_2$.
The spectrum depends only on the angles $\Phi=l\phi+m\psi$,
$\Psi=-m\phi+(l+m)\psi$
attached to the translations
${\bf T}_1$, ${\bf T}_2$.
In the $N=3p$ case, the absolute minimum of the energy is obtained for
angles $\Phi$ ,$\Psi$ equal to
0 or $\pm\frac{2\pi}{3}$ characterizing
the classical N\'eel states
and only $(\Phi,\Psi)=(\pm\frac{2\pi}{3},\mp\frac{2\pi}{3})$
in the $N=3p+1$ samples.}
\label{fig:twist-ener}
\end{figure}

\begin{figure}
\caption[99]{Spectra of $N=3p+1$ samples versus $ S_{3}^{2} $.
Note the tower of states, only doubly degenerate, and its
collapse to the absolute ground-state with increasing $N$.}
\label{fig:en-nonmult}
\end{figure}

\begin{figure}
\caption[99]{Finite-size scaling of the ground-state energy per bond and
comparison with spin-wave results.

$a)$ three-legs symbols: diagonalization results for the $N=3p$-samples;
black triangles: averaged values computed thanks to Eq.18;
small open triangles: spin-wave results;
dotted line: $N^{-3/2}$ fit on large $N$ spin-wave results.

$b)$ same as in $a)$ except crosses: diagonalization results for
the $N=3p+1$-samples.

$c)$ averaged values of $ < 2 {\bf s}_{i}.{\bf s}_{j}> $
versus $N^{-3/2}$; dotted line : $N^{-3/2}$ fit
to these results of diagonalizations (see text).}
\label{fig:finit-en}
\end{figure}

\begin{figure}
\caption[99]{Static structure function in the ground-state
of the $N=3p$ samples versus the ${\bf k}$-vector modulus.}
\label{fig:static-fact}
\end{figure}

\begin{figure}
\caption[99]{N\'eel order parameter as a function of the sample
size. The order parameter is normalized by its maximum value
(see Eq.\ref{eq-order} and Eq.\ref{eq-order1}).
Triangles (resp. squares) stand for $N=3p$- (resp. $N=3p+1$)-samples;
black symbols stand for diagonalization results;
open symbols stand for first order spin-wave results
($2 < M_{z} > $ of Eq.\ref{eq-SWOP});
dotted lines :
large $N$ fits ($ N > 5000 $) of first-order spin-wave results
($X_{N}=X_{\infty} + a N^{-1/2}$).
}
\label{fig:QDJ-OP}
\end{figure}

\end{document}